\documentclass[10pt, conference, compsocconf]{IEEEtran}

\usepackage{picins}

\usepackage{times}
\usepackage{helvet}
\usepackage{courier}

\usepackage{algorithm}
\usepackage{multirow,tabularx}
\usepackage{balance}

\usepackage{epsfig}
\usepackage{subfigure}
\usepackage{amsmath,amssymb}
\usepackage{slashbox}

\usepackage{times}
\usepackage{amsmath}
\usepackage{amssymb}
\usepackage{multirow}
\usepackage{graphicx}
\usepackage{wrapfig}
\usepackage{subfig}
\usepackage{url}
\usepackage{enumerate}

\usepackage{latexsym}
\usepackage{amsmath,amssymb}
\usepackage{multirow}
\usepackage{url}
\usepackage{epsfig}
\usepackage{subfigure}

\newcommand{\B}{\mathbf{B}}

\newcommand{\I}{\mathbf{I}}

\newcommand{\R}{\mathbf{R}}

\newcommand{\U}{\mathbf{U}}
\newcommand{\V}{\mathbf{V}}

\newcommand{\X}{\mathbf{X}}
\newcommand{\Y}{\mathbf{Y}}

\newcommand{\x}{\boldsymbol{x}}

\newcommand{\w}{\boldsymbol{w}}

\newcommand{\br}{\boldsymbol{r}}

\newcommand{\ignore}[1]{}

\begin{document}

\title{Collaborative Recommendation with Auxiliary Data: A Transfer Learning View}

% =========================================================================
\author{\IEEEauthorblockN{Weike Pan}
\IEEEauthorblockA{
College of Computer Science and Software Engineering, Shenzhen University\\
panweike@szu.edu.cn}
}

% =========================================================================================

\maketitle
\begin{abstract}
Intelligent recommendation technology has been playing an increasingly important role in various industry applications such as e-commerce product promotion and Internet advertisement display. Besides users' feedbacks (e.g., numerical ratings) on items as usually exploited by some typical recommendation algorithms, there are often some additional data such as users' social circles and other behaviors. Such auxiliary data are usually related to users' preferences on items behind the numerical ratings. Collaborative recommendation with auxiliary data (CRAD) aims to leverage such additional information so as to improve the personalization services, which have received much attention from both researchers and practitioners.

Transfer learning (TL) is proposed to extract and transfer knowledge from some auxiliary data in order to assist the learning task on some target data. In this paper, we consider the CRAD problem from a transfer learning view, especially on how to achieve knowledge transfer from some auxiliary data. First, we give a formal definition of transfer learning for CRAD (TL-CRAD). Second, we extend the existing categorization of TL techniques (i.e., adaptive, collective and integrative knowledge transfer algorithm styles) with three knowledge transfer strategies (i.e., prediction rule, regularization and constraint). Third, we propose a novel generic knowledge transfer framework for TL-CRAD. Fourth, we describe some representative works of each specific knowledge transfer strategy of each algorithm style in detail, which are expected to inspire further works. Finally, we conclude the paper with some summary discussions and several future directions.
\end{abstract}


% =========================================================================================
\section{Introduction}
\label{sec:intr}

Intelligent recommendation technology has been a standard component in most Internet systems such as e-commerce and advertisement systems, which is embedded in order to provide personalization services. There are two main approaches widely used in personalized recommendation for an active user~\cite{Next-RS-TKDE05,RS-introduction-book-10,RS-handbook-10}, i.e., content-based recommendation and collaborative recommendation. Content-based methods promote an item based on the relevance between a candidate item and the active user's consumed items, while collaborative recommendation techniques focus on collective intelligence and exploit the community's data so as to recommend preferred items from similar-taste users. However, both methods are limited to users' feedbacks of explicit scores or implicit examinations, which may result in the sparsity problem due to the lack of users' behaviors.

Fortunately, there are often some additionally available data besides the users' feedbacks (e.g., numerical ratings) in a recommendation system. There are at least four types of auxiliary data as shown in Table~\ref{tbl:Auxiliary-Data}, such as content information~\cite{CMF-Singh-KDD08}, time contextual information~\cite{KDD09-Koren-Temporal-Dynamics}, social networks~\cite{UMUAI2013-Facebook} and additional feedbacks~\cite{Nathan-Feedback-CIKM2010}. These auxiliary data have the potential to help reduce the aforementioned sparsity effect and thus improve the recommendation performance. In this paper, we study on how to exploit different types of auxiliary data in collaborative recommendation, which is coined as {\em collaborative recommendation with auxiliary data} (CRAD).

% --------------------------------------------------------------
\begin{table}[htbp]
\caption{A list of auxiliary data.} \label{tbl:Auxiliary-Data}
\begin{center}

\begin{tabular}{ l|l} \hline\hline

% ---------------
\multicolumn{2}{l}{\bf Content} \\ \hline
& user's static profile of demographics, affiliations, etc.\\ \cline{2-2}

& item's static description of price, brand, location, etc. \\ \cline{2-2}

& user-item pair's user generated content (UGC), etc. \\ \hline

% ---------------
\multicolumn{2}{l}{\bf Context} \\ \hline
& user's dynamic context of mood, health, etc.\\ \cline{2-2}

& item's dynamic context of remaining quantities, etc.\\ \cline{2-2}

& user-item pair's dynamic context of time, etc.\\ \hline

% ---------------
\multicolumn{2}{l}{\bf Network} \\ \hline
& user-user social network of friendship, etc.\\ \cline{2-2}

& item-item relevance network of taxonomy, etc.\\ \cline{2-2}

& user-item-user network of sharing items with friends, etc.\\ \hline

% ---------------
\multicolumn{2}{l}{\bf Feedback} \\ \hline
& user's  feedback of rating on other items, etc.\\ \cline{2-2}

& item's  feedback of browsing by other users, etc.\\ \cline{2-2}

& user-item pair's feedback of collection, etc.\\ \hline \hline

\end{tabular}
\end{center}
\end{table}
% --------------------------------------------------------------

Specifically, we study the CRAD problem from a {\em transfer learning}~\cite{Sinno-TL-suvery} view, in which we consider the users' feedback data as our {\em target data}, and all other additional information as our {\em auxiliary data}. In particular, we focus on how to achieve knowledge transfer from some auxiliary data to a target data, aiming to answer the fundamental question of transfer learning~\cite{Sinno-TL-suvery}, i.e., ``how to transfer''. With this focus, we extend previous categorization of transfer learning techniques in collaborative filtering~\cite{Weike-Pan-IJCAI11,AAAI2012-TIF}, and answer the above question from two dimensions, including {\em knowledge transfer algorithm styles} (i.e., adaptive, collective and integrative knowledge transfer) and {\em knowledge transfer strategies} (i.e., prediction rule, regularization and constraint). We then propose a novel generic knowledge transfer framework and describe some representative works if any in each category to answer the ``how to transfer'' question in detail, in particular of the main idea that may be generalized to other applications. Finally, we conclude the paper with some summary discussions and several exciting future directions.

% =========================================================================================
\section{Collaborative Recommendation with Auxiliary Data}

% =========================================================================================
\subsection{Problem Definition}

\begin{figure}[!htb]
\begin{center}
\psfig{figure=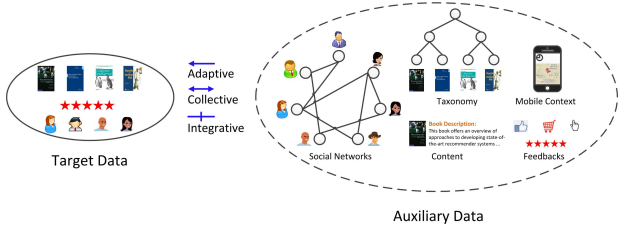,width=3.2in}
\end{center}
\caption{Illustration of Transfer Learning for Collaborative Recommendation with Auxiliary Data (TL-CRAD).}
\label{fig:TL-CRAD}
\end{figure}

We have a target data and an auxiliary data. In the target data, we have some feedbacks from $n$ users and $m$ items, which is usually represented as a rating matrix $\R = [r_{ui}]^{n\times m}$ and an indicator matrix $\Y \in \{0, 1\}^{n\times m}$, where $y_{ui}=1$ means that the feedback $r_{ui}$ is observed. In the auxiliary data, we have some additional data such as content, context, network and feedback information as shown in Table~\ref{tbl:Auxiliary-Data}. Our goal is then to predict the unobserved feedbacks in $\R$ by transferring knowledge from the available auxiliary data. We illustrate the studied problem in Figure~\ref{fig:TL-CRAD}, where the left part is the target data of users' feedbacks and the right part denotes different types of auxiliary data.

% -----------------------------------------------------------------------------------------
% -----------------------------------------------------------------------------------------

% =========================================================================================
\subsection{Categorization of Knowledge Transfer}

Following the fundamental question of ``how to transfer'' in transfer learning~\cite{Sinno-TL-suvery,AAAI2012-TIF}, we first categorize various transfer learning algorithms into adaptive knowledge transfer, collective knowledge transfer and integrative knowledge transfer w.r.t. {\em knowledge transfer algorithm styles}. For each type of algorithm styles, we then study the related works in three specific {\em knowledge transfer strategies}, including transfer via prediction rule, transfer via regularization and transfer via constraint, which are closely related to the three parts of a typical optimization problem~\cite{Book-Boyd-2004}, i.e., loss function, regularization and constraint.

Note that the binary categorization of adaptive knowledge transfer and collective knowledge transfer was first briefly described in~\cite{Weike-Pan-IJCAI11}, and was later expanded with one more category of integrative knowledge transfer in~\cite{AAAI2012-TIF}. And in this paper, we further expand it with three specific knowledge transfer strategies in each algorithm style.

% =========================================================================================
\subsection{A Generic Knowledge Transfer Framework}

In this paper, we mainly focus on some recent works of low-rank transfer learning methods for collaborative recommendation with auxiliary data (CRAD), in particular of matrix factorization based methods. The prosperity of matrix factorization based methods are mostly due to many successful stories in various public competitions and reported industry applications. Factorization based methods are also the state-of-the-art in TL-CRAD because they are able to digest the sparse rating data well via learning latent variables and are also flexible to incorporate different types of auxiliary data.

Mathematically, matrix factorization based methods can be formulated by a loss function and a regularization term, i.e., $\min_{\Theta} \mathcal{E} ( \Theta | \R) + \mathcal{R}(\Theta)$, where $\Theta$ is the model parameter. We extend such basic formulation and propose a novel generic framework for TL-CRAD,
\begin{equation}\label{eq:TL-CRAD-framework}
\begin{array}{l}
    \min_{\Theta, \mathbb{K}} \; \mathcal{E} ( \Theta, \mathbb{K} | \R, \mathbb{A}) + \mathcal{R}(\Theta | \mathbb{K}, \mathbb{A}) + \mathcal{R}(\mathbb{K})  \\
    \mbox{s.t.} \; \Theta \in \mathcal{C}(\mathbb{K}, \mathbb{A})
\end{array}
\end{equation}
which contains a loss function $\mathcal{E} ( \Theta, \mathbb{K} | \R,  \mathbb{A})$, two regularization terms $\mathcal{R}(\Theta | \mathbb{K}, \mathbb{A})$ and $\mathcal{R}(\mathbb{K})$, and a
constraint $\Theta \in \mathcal{C}(\mathbb{K}, \mathbb{A})$.
Specifically, $\R$ is the target user-item rating matrix, $\mathbb{A}$ is the auxiliary data, $\mathbb{K}$ is the extracted knowledge from $\mathbb{A}$, and $\Theta$ is the model parameter. Note that the prediction rule is not explicitly shown but embedded in the loss function $\mathcal{E} ( \Theta, \mathbb{K} | \R, \mathbb{A})$. In the following sections, we will describe some representative works of TL-CRAD, which are instantiations of the generic framework in Eq.(\ref{eq:TL-CRAD-framework}).

% =========================================================================================
\section{Adaptive Knowledge Transfer}
\label{sec:adaptive}

Adaptive knowledge transfer aims to {\em adapt} the knowledge, i.e., $\mathbb{K}$, extracted from an auxiliary data to a target data, which is a {\em directed} knowledge transfer approach similar to traditional domain adaptation methods. In this section, we describe two adaptive knowledge transfer strategies as instantiated from Eq.(\ref{eq:TL-CRAD-framework}), including
(i) transfer via regularization, $\min_{\Theta} \mathcal{E} ( \Theta | \R) + \mathcal{R}(\Theta | \mathbb{K})$, and (ii) transfer via constraint, $\min_{\Theta} \mathcal{E} ( \Theta | \R), \mbox{s.t.} \; \Theta \in \mathcal{C}(\mathbb{K})$.

% =========================================================================================
\subsection{Transfer via Regularization}

{\noindent \bf CST (Coordinate System Transfer)} \; CST~\cite{Weike-Pan-AAAI10} studies knowledge transfer from auxiliary implicit feedbacks of browsing records to target explicit feedbacks of ratings. Specifically, it incorporates the coordinate systems (or latent features) extracted from auxiliary data into the target factorization system (i.e., $\Y \! \odot \! \R \! \sim \! \U \B \V^T$) via two biased regularization terms~\cite{Weike-Pan-AAAI10},
\begin{eqnarray}\label{eq:CST}
||\U-\dot{\U}||_F^2 + ||\V-\ddot{\V}||_F^2,
\end{eqnarray}
where $\dot{\U} \in \mathbb{R}^{n\times d}$ and $\ddot{\V} \in \mathbb{R}^{m \times d}$ are user-specific feature matrix and item-specific feature matrix, respectively. The two biased regularization terms in Eq.(\ref{eq:CST}) are used to constrain the latent feature matrices $\U \in \mathbb{R}^{n\times d}$ and $\V \in \mathbb{R}^{m\times d}$ to be similar to $\dot{\U}$ and $\ddot{\V}$, respectively. Note that the factorization system $\Y \! \odot \! \R \! \sim \! \U \B \V^T$ denotes the approximation of $\Y \! \odot \!\R$ via matrix tri-factorization $\U \B \V^T$, where $\U$ and $\V$ are orthonormal matrices (i.e., $\U^T \U=\I$, $\V^T \V = \I$). Empirically, CST~\cite{Weike-Pan-AAAI10} works well when the auxiliary data are dense while the target ratings are few.

Biased regularization is a classical approach commonly used in machine learning, in particular of domain adaptation methods~\cite{biasedSVM-ICML06}. Note that it may also be considered as a soft constraint as compared with the hard constraint in collective knowledge transfer methods~\cite{Weike-Pan-IJCAI11,CMF-Singh-KDD08} in Section~\ref{sec:collective-constraint}.

% =========================================================================================
\subsection{Transfer via Constraint}

{\noindent \bf CBT (CodeBook Transfer)} \; CBT~\cite{Bin-Li-IJCAI09} is an early transfer learning algorithm, which studies knowledge transferability between two distinct data, which may be regarded as {\em far transfer of learning} in psychology~\cite{Book2001-TransferOfLearning}. Specifically, it transfers knowledge of cluster-level rating behavior from auxiliary data of movies to target data of books. First, a cluster-level rating pattern (a.k.a., codebook), $\breve{\B} \in \mathbb{R}^{d\times d}$, is constructed from the auxiliary data $\breve{\R} \in \mathbb{R}^{\breve{n} \times \breve{m}}$ via some co-clustering algorithm, where each entry of $\breve{\B}$ denotes the average rating of the corresponding co-cluster. Second, the codebook is transferred to the target data via codebook expansion $\U \B \V^T$ with the following constraint~\cite{Bin-Li-IJCAI09},
\begin{eqnarray}
\B=\breve{\B},
\end{eqnarray}
which means that the rating pattern is shared between target data and auxiliary data. Note that $\U \in \{0,1\}^{n\times d}$ and $V \in \{0,1\}^{m\times d}$ are membership indicator matrices.

A later extension called RMGM~\cite{Bin-Li-ICML09} combines codebook construction and codebook expansion in CBT~\cite{Bin-Li-IJCAI09} into one single step with soft membership indicator matrices. Considering the existence of more than one auxiliary data, the codebook in CBT~\cite{Bin-Li-IJCAI09} may also be extended to multiple codebooks with different relatedness weight~\cite{CIKM2012-TALMUD}. Furthermore, a recent work generalizes the codebook to include a data-independent rating pattern and a data-dependent rating pattern, which is shown to be more accurate than sharing the data-independent common knowledge only~\cite{ECMLPKDD2013-CrossDomainRecommendation}.

The idea of transferring compact group-level knowledge may also be applied to other applications such as text mining and bio-informatics. For example, MTrick~\cite{FuzhenZHUANG-codebook-SDM10} extends CBT~\cite{Bin-Li-IJCAI09} and RMGM~\cite{Bin-Li-ICML09} with supervised label information and studies its effectiveness in cross-domain document categorization. Furthermore, the $2$-D cluster-level rating pattern may also be generalized to high dimensions, such as transferring a $3$-D cluster of knowledge from an auxiliary tagging data in the form of (user, item, tag) to a target one~\cite{CrossDomain3D-KDD2013}.

Cluster-level rating pattern in the above works is a kind of group behavior, which is more stable than individual behavior and has higher transferability. It is thus particularly useful when the explicit correspondences or overlaps are not available between entities of target data and auxiliary data.

% =========================================================================================
\section{Collective Knowledge Transfer}

Collective knowledge transfer usually {\em jointly} learns the shared knowledge $\mathbb{K}$ and not-shared effect of each data simultaneously, which is a {\em bi-directed} knowledge transfer approach with richer interactions similar to multi-task learning algorithms. We describe some representative works of collective knowledge transfer via constraint on model parameters, $\min_{\Theta, \mathbb{K}} \mathcal{E}( \Theta | \R) + \mathcal{R}(\Theta) + \mathcal{E}( \mathbb{K} | \mathbb{A}) + \mathcal{R}(\mathbb{K}),  \mbox{s.t.} \; \Theta \in \mathcal{C}(\mathbb{K})$, which is also an instantiation of Eq.(1). Note that the model parameter $\Theta$ and shared knowledge $\mathbb{K}$ are learned simultaneously, instead of in two separate steps as that in adaptive knowledge transfer.

% =========================================================================================
\subsection{Transfer via Constraint}
\label{sec:collective-constraint}

\vspace{0.05in}
{\noindent \bf CMF (Collective Matrix Factorization)} \; CMF~\cite{CMF-Singh-KDD08} is proposed to collectively factorize one user-item rating matrix $\R \in \mathbb{R}^{n\times m}$, $\Y \! \odot \! \R \! \sim \! \U \V^T$, and one item-content matrix $\ddot{\R} \in \mathbb{R}^{\ddot{n} \times m}$, $\ddot{\R} \! \sim \! \ddot{\U} \ddot{\V}^T$, with the idea of sharing the same item-specific latent features $\V$~\cite{CMF-Singh-KDD08},
\begin{eqnarray}
    \V = \ddot{\V},
\end{eqnarray}
which means that the item-specific latent feature matrix $\ddot{\V}$ is shared as a bridge to enable knowledge transfer between two data. We may also use different link functions on the factorized variables $f(U_{u\cdot} V_{i\cdot}^T)$~\cite{CMF-Singh-KDD08}. A similar model is proposed independently in the context of social recommendation~\cite{CIKM08-Hao-Ma-SoRec}, which generalizes the basic matrix factorization model by jointly factorizing a user-item rating matrix $\R \in \mathbb{R}^{n\times m}$, $\Y \! \odot \! \R \! \sim \! \U \V^T$, and a user-user social network matrix $\dot{\R} \in \mathbb{R}^{n\times n}$,  $\dot{\R} \! \sim \! \dot{\U} \dot{\V}$, with the constraint of $\U = \dot{\U}$.

The underlying assumption that same users (or items) in target data and auxiliary data are associated with the same latent factors is quite universal. Various models with a similar spirit have been proposed to fuse user-side and/or item-side auxiliary data via sharing latent features or topic distributions. WNMCTF~\cite{Jiho-WNMCTF-ACML09} follows non-negative matrix tri-factorization (NMF)~\cite{Lee-NIPS-2001} and collectively factorizes one user-item rating matrix, one user-demographics matrix and one item-content matrix with the constraint of sharing both the same user-specific latent feature matrix and item-specific latent feature matrix. MRMF~\cite{Lippert-MRMF-NIPS08-SISOworkshop} and HYRES~\cite{CIKM09-TSA-Workshop-Jakob-MRMF} collectively factorize more than two matrices from both user-side and item-side content information. JMF~\cite{Shi-JMF-TIST13} collectively factorizes one user-item rating matrix and one item-item similarity matrix mined from item-side auxiliary data of movies' mood descriptions. MCF-LF~\cite{YuZhang-UAI-2010}, CLP-GP~\cite{Bin-ICML2010} and NB-MCF~\cite{CIKM2013-Nonparametric} study multiple user-side auxiliary data matrices and learn users' preferences and similarities between target and auxiliary data simultaneously, which are shown to be more effective as compared with sharing the latent features alone.
LOCABAL~\cite{IJCAI2013-JiliangTANG-LOCABAL} collectively factorizes one user-item rating matrix weighted by users' global reputations and one user-user social matrix weighted by consine similarities, with the constraint of sharing the same user-specific latent feature matrix. STLCF~\cite{SelectiveTLCF-SDM2013} learns the user preferences in a joint and selective manner from multiple user-aligned data via selectively transferring high quality knowledge of consistent data, which is empirically more accurate than collective knowledge transfer without selection.

TCF~\cite{Weike-Pan-IJCAI11} collectively factorizes a 5-star numerical target data $\R$ and a binary like/dislike auxiliary data, and assumes that both user-specific and item-specific latent feature matrices are the same. Besides the shared latent features, TCF uses two inner matrices to capture the data-dependent information, which is different from the inner matrix used in CBT~\cite{Bin-Li-IJCAI09} and RMGM~\cite{Bin-Li-ICML09}. The strategy to share some common knowledge and to not share some specific effect is a sophisticated knowledge transfer approach, which is expected to be more applicable to other applications. A recent extension called interaction-rich TCF (iTCF)~\cite{IEEE-IS2014-iTCF} achieves a good balance between the efficiency of stochastic update rules or stochastic learning in CMF~\cite{CMF-Singh-KDD08} and the effectiveness of knowledge transfer for heterogeneous data in TCF~\cite{Weike-Pan-IJCAI11}. iTCF is an efficient transfer learning algorithm, which shares user-specific latent features in a smooth manner besides requiring the same item-specific latent features.

% =========================================================================================
\section{Integrative Knowledge Transfer}

Integrative knowledge transfer {\em incorporates} the raw auxiliary data, i.e., $\mathbb{A}$, as known knowledge into the target learning task via the prediction rule, regularization terms or additional constraints, which can be considered as an {\em embedded} knowledge transfer approach similar to feature engineering, information fusion and data integration methods. Mathematically, we can instantiate the generic framework in Eq.(\ref{eq:TL-CRAD-framework}), and have (i) transfer via prediction rule, $\min_{\Theta} \mathcal{E} ( \Theta | \R, \mathbb{A}) + \mathcal{R}(\Theta)$, (ii) transfer via regularization, $\min_{\Theta} \mathcal{E} ( \Theta | \R ) + \mathcal{R}(\Theta | \mathbb{A})$, and (iii) transfer via constraint, $\min_{\Theta} \mathcal{E} ( \Theta | \R) + \mathcal{R}(\Theta), \mbox{s.t.} \; \Theta \in \mathcal{C}(\mathbb{A})$. It is interesting to see that we include the raw auxiliary data $\mathbb{A}$ instead of the extracted knowledge $\mathbb{K}$, which is thus different from that of adaptive knowledge transfer as shown in Section~\ref{sec:adaptive}.

% =========================================================================================
\subsection{Transfer via Prediction Rule}

Typically, once a recommendation model has been built using some training data, we can use a prediction rule such as~\cite{KDD09-Koren-Temporal-Dynamics} $\hat{r}_{ui} = \mu + b_u + b_i + U_{u\cdot} V_{i\cdot}^T$ or $\hat{r}_{ui} = U_{u\cdot} V_{i\cdot}^T$ in order to predict user $u$'s preference on item $i$. Note that $U_{u\cdot}, V_{i\cdot} \in \mathbb{R}^{1\times d}$ are user $u$'s and item $i$'s latent feature vectors, respectively, and $\mu$ is the global mean, $b_u$ is user $u$'s bias, and $b_i$ is item $i$'s bias.

\vspace{0.05in}
{\noindent \bf FM (Factorization Machines)} \; FM~\cite{TIST12-libFM} represents the user-item feedback matrix $\R$ in a novel way, i.e., a design matrix $\X \in \{1,0\}^{q \times (n+m)}$ and a rating vector $\br \in \{1,2,3,4,5\}^{q\times 1}$, where $q$ is the number of ratings in $\R$. For a rating triple $(u, i, r_{ui})$ of the user-item feedback matrix $\R$, the corresponding row of the design matrix is $\x = \{ (u,1), (i,1) \} \in \mathbb{R}^{1\times (n+m)}$, where the $u$th and $(n+i)$th entries are $1$s and all other entries are $0$s, and the value of the corresponding entry of the rating vector $\br$ is $r_{ui}$. We then have a revised prediction rule with pairwise interactions between latent factors~\cite{TIST12-libFM},
\begin{eqnarray}\label{eq:FM-prediction-rule}
  \hat{r}_{ui} = w_0 + \sum_{j=1}^{n+m} w_j x_j + \sum_{j=1}^{n+m} \sum_{j'=j+1}^{n+m} x_j x_{j'} w_{jj'}
\end{eqnarray}
where $w_{jj'}$ denotes the inner product of two latent feature vectors. With the new representation via the design matrix, we can augment it with some auxiliary data in a simple but effective pre-processing step of feature engineering, such as user-side auxiliary ratings~\cite{ECIR2014-cross-domain-CF-FM}. Note that when no auxiliary data is fused, the prediction rule is the same as that of basic matrix factorization, i.e., $\hat{r}_{ui} = w_0 + w_u + w_{n+i} + w_{u, n+i}$, where $w_0$ is the global mean (i.e., $\mu$), $w_u$ is user $u$'s bias (i.e., $b_u$), $w_{n+i}$ is item $i$'s bias (i.e., $b_i$), and $w_{u, n+i} = U_{u\cdot} V_{i\cdot}^T$ is the inner product of user $u$'s and item $i$'s latent feature vectors.

Besides using the design matrix in FM~\cite{TIST12-libFM}, there are some other approaches to incorporate auxiliary data via a revised prediction rule. RSTE~\cite{HaoMa-RSTE-TIST11} designs a mixed prediction rule with two terms, $\hat{r}_{ui} = \lambda U_{u\cdot} V_{i\cdot}^T + (1-\lambda) \sum_{u' \in T_{u}^+} \tilde{e}_{u'i} U_{u'\cdot}V_{i\cdot}^T$, where $T^{+}_{u}$ is the set of trusted friends of user $u$ and $\sum_{u' \in T_{u}^+} \tilde{e}_{u'i} U_{u'\cdot}V_{i\cdot}^T$ represents the friends' overall taste on item $i$. Note that $\tilde{e}_{u'i}$ is estimated from the user-side social networks or the target user-item rating matrix~\cite{HaoMa-RSTE-TIST11}. BMPSI~\cite{Porteous-BPMF-side-infor-AAAI10} designs an integrated prediction rule for both target feedback data and auxiliary data, $\hat{r}_{ui} = U_{u\cdot} V_{i\cdot}^T + \w_u \dot{\x}_{ui}^T + \w_i \ddot{\x}_{ui}^T$, where $\dot{\x}_{ui} \in \mathbb{R}^{1\times \dot{d}_x}$ and $\ddot{\x}_{ui} \in \mathbb{R}^{1\times \ddot{d}_x}$ are user-side and item-side raw features related to the rating at $(u,i)$ of $\R$, including the rating's time information, user $u$'s latest two ratings, the ratings the user $u$ gave on those $5$ most similar movies measured by Pearson correlation coefficient, the features of movie directors and actors extracted from Wikipedia\footnote{http://www.wikipedia.org/}, etc. Another recent feature engineering based model called SVDfeature~\cite{JMLR2012-SVDfeature} is an efficient but restricted case of FM, which is also able to incorporate users' demographics, items' descriptions and contextual information. The rich pairwise interactions in FM~\cite{TIST12-libFM} as shown in Eq.(\ref{eq:FM-prediction-rule}) are able to capture more complex correlations among the variables, which are thus expected to generate better recommendations.

Integrating auxiliary data into the prediction rule is an effective approach for knowledge transfer, where the knowledge of the raw auxiliary data (or part of model parameters more specifically) is learned automatically. However, the revised prediction rule will also cause the learning and prediction procedures more expensive regarding the time and space complexity.

% =========================================================================================
\subsection{Transfer via Regularization}

The main idea of integrative transfer via regularization is to constrain the latent feature matrices or vectors to be similar between related users or items, e.g., similar users according to tagging information~\cite{Yi-Zhen-tagiCoFi-RecSys09} or socially connected users~\cite{Mohsen-SocialNetwork-RecSys10}.

\vspace{0.05in}
{\noindent \bf TagiCoFi (Tag Informed Collaborative Filtering)} \; TagiCoFi~\cite{Yi-Zhen-tagiCoFi-RecSys09} is proposed for incorporating social tagging data into target numerical rating data. Specifically, it first obtains a user-user similarity matrix from social tagging data and then introduces an additional regularization term to the basic matrix factorization~\cite{Yi-Zhen-tagiCoFi-RecSys09},
\begin{eqnarray}\label{eq:regularization-TagiCoFi}
\sum_{u=1}^{n} \sum_{u'=1}^n \dot{s}_{uu'}|| U_{u\cdot} - U_{u'\cdot} ||_F^2,
\end{eqnarray}
where $\dot{s}_{uu'}$ is the similarity between users $u$ and $u'$, and thus transfers knowledge of the nearest neighbors' taste via constraining the user-specific features to be similar in the latent space.

SocialMF~\cite{Mohsen-SocialNetwork-RecSys10} studies the effect of trust propagation and generalizes the basic matrix factorization model by introducing a different additional regularization term from trusted friends~\cite{Mohsen-SocialNetwork-RecSys10},
\begin{eqnarray}\label{eq:regularization-SocialMF}
 \sum_{u=1}^{n} || U_{u\cdot} - \sum_{u' \in T_{u}^+} \dot{s}_{uu'} U_{u'\cdot} ||_F^2
\end{eqnarray}
where $T_{u}^+$ is the set of trusted friends of user $u$ (not including user $u$ himself/herself) and $\dot{s}_{uu'}$ is the similarity between users $u$ and $u'$ obtained from social networks. Similarly, it transfers knowledge of the friends' taste via constraining the user-specific features to be similar in the latent space.

We can see that the regularization term in Eq.(\ref{eq:regularization-TagiCoFi}) focus on the distance between a user' feature vector and each of his/her friends' feature vectors, while the regularization term in Eq.(\ref{eq:regularization-SocialMF}) defines the distance between one user's feature vector and a weighted sum of his/her friends' feature vectors. A recent work on TV channel recommendation studies the effect of combining those two regularization terms for both users and items and obtains better recommendation performance~\cite{KBS2013-HybridRegularizationTerms}, which shows the complementary effect of those two types of regularization in knowledge transfer. Another interesting point is that the regularization terms in Eq.(\ref{eq:regularization-TagiCoFi}), Eq.(\ref{eq:regularization-SocialMF}) and Eq.(\ref{eq:CST}) can all be considered as soft constraint imposed on latent features.

Knowledge transfer via regularization will usually increase the computational complexity in the training step, but remains the same in the prediction step since the prediction rule is not changed, which is thus more efficient than the aforementioned strategy of transfer via prediction rule.

% =========================================================================================
\subsection{Transfer via Constraint}

{\noindent \bf TIF (Transfer by Integrative Factorization)} \; TIF~\cite{AAAI2012-TIF} aims to incorporate knowledge from social impressions or anchoring effects, which are represented as score intervals called uncertain ratings. Specifically, it studies on how to leverage auxiliary uncertain ratings, denoted $[a_{ui}, b_{ui}]$s, to the target data of numerical ratings. Different from most previous works, it incorporates auxiliary data through some constraints defined on the score intervals in addition to the basic matrix factorization~\cite{AAAI2012-TIF},
\begin{eqnarray}
    \hat{r}_{ui} \in \mathcal{C}(a_{ui}, b_{ui})
\end{eqnarray}
where the constraint, $\hat{r}_{ui} \in \mathcal{C}(a_{ui}, b_{ui})$, requires that the estimated preference by the learned model should be in the range of the corresponding auxiliary uncertain rating.

Integrative knowledge transfer via constraint is related to knowledge-based recommendation~\cite{RS-introduction-book-10}, where a user's additional constraints or requirements need to be satisfied during recommendation. Incorporating auxiliary data via constraints are also flexible since auxiliary data can usually be represented as some constraints.

% =========================================================================================
\section{Discussions and Future Directions}

% =========================================================================================
\subsection{Discussions}
\label{sec:conclusion}

% -----------------------------------------------------------------------------------------
\begin{table*}[htbp]
\caption{Some representative works of transfer learning for collaborative recommendation with auxiliary data (TL-CRAD) in the perspective of ``how to transfer'' in transfer learning, including different knowledge transfer algorithm styles and different knowledge transfer strategies. We also include the corresponding mathematical formulations as instantiations of the generic framework in Eq.(\ref{eq:TL-CRAD-framework}), i.e., $\min_{\Theta, \mathbb{K}} \mathcal{E} ( \Theta, \mathbb{K} | \R, \mathbb{A}) + \mathcal{R}(\Theta | \mathbb{K}, \mathbb{A}) + \mathcal{R}(\mathbb{K}), \mbox{s.t.} \Theta \in \mathcal{C}(\mathbb{K}, \mathbb{A})$.} \label{tbl:summary-TL-CRAD-techniques}
\begin{center}

\begin{tabular}{  l | l | l | l } \hline\hline

\backslashbox{Style}{Strategy}
& Prediction rule & Regularization & Constraint\\
\hline \hline

\multirow{2}{*}{Adaptive} & & $\min_{\Theta} \mathcal{E} ( \Theta | \R) + \underline{\mathcal{R}(\Theta | \mathbb{K})}$ & $\min_{\Theta} \mathcal{E} ( \Theta | \R), \; \underline{\mbox{s.t.} \; \Theta \in \mathcal{C}(\mathbb{K})}$ \\
& & e.g., CST{~\cite{Weike-Pan-AAAI10}} & e.g., CBT{~\cite{Bin-Li-IJCAI09}} \\ \hline

\multirow{3}{*}{Collective} & & & $\min_{\Theta, \mathbb{K}} \mathcal{E}( \Theta | \R) + \mathcal{R}(\Theta)$ \\
& & & \;\;\;\;\;\; $+\mathcal{E}( \mathbb{K} | \mathbb{A}) + \mathcal{R}(\mathbb{K}),  \; \underline{\mbox{s.t.} \; \Theta \in \mathcal{C}(\mathbb{K})}$ \\
 & & & e.g., CMF{~\cite{CMF-Singh-KDD08}}, RMGM{~\cite{Bin-Li-ICML09}}, TCF{~\cite{Weike-Pan-IJCAI11}} \\ \hline

\multirow{2}{*}{Integrative} & $\min_{\Theta} \underline{\mathcal{E} ( \Theta | \R, \mathbb{A})} + \mathcal{R}(\Theta)$ & $\min_{\Theta} \mathcal{E} ( \Theta | \R ) + \underline{\mathcal{R}(\Theta | \mathbb{A})}$ & $\min_{\Theta} \mathcal{E} ( \Theta | \R) + \mathcal{R}(\Theta),  \; \underline{\mbox{s.t.} \; \Theta \in \mathcal{C}(\mathbb{A})}$ \\
 & e.g., FM{~\cite{TIST12-libFM}}, BMFSI{~\cite{Porteous-BPMF-side-infor-AAAI10}} & e.g., tagiCoFi{~\cite{Yi-Zhen-tagiCoFi-RecSys09}} & e.g., TIF{~\cite{AAAI2012-TIF}}\\

\hline \hline

\end{tabular}
\end{center}
\end{table*}
% -----------------------------------------------------------------------------------------

We summarize some representative works of transfer learning for collaborative recommendation with auxiliary data (TL-CRAD) in Table~\ref{tbl:summary-TL-CRAD-techniques}. We can see that integrative knowledge transfer via prediction rule and collective knowledge transfer via constraint have recently received more attention, which are also the state-of-the-art TL-CRAD algorithms w.r.t. recommendation accuracy in corresponding problem settings. The interaction between auxiliary data and target data usually becomes richer from adaptive, collective to integrative algorithm styles, which are believed to enable more effective knowledge transfer. However, the time complexity may also increase from adaptive to integrative algorithm styles, especially of the cost caused by sophisticated prediction rules and regularization terms used in integrative knowledge transfer approaches. We can also see that there are some blank and few-work entries in Table~\ref{tbl:summary-TL-CRAD-techniques}, which provide opportunities for further studies.

Parallel to various {\em data modeling} methods in Table~\ref{tbl:summary-TL-CRAD-techniques}, there are also some recommendation approaches based on {\em user modeling}, which may be developed for TL-CRAD so as to further expand the two-dimension categorization used in Table~\ref{tbl:summary-TL-CRAD-techniques}. A recent brief survey~\cite{ICTAI2011-BinLI-CrossDomainCF} studies cross-domain collaborative filtering in the perspective of {\em collaborative domain} (i.e., source domain and target domain in classic transfer learning~\cite{Sinno-TL-suvery}) and {\em knowledge transfer style}, which is different from our focus on ``how to transfer'' in transfer learning in a more general and practical recommendation problem. An extended survey of cross-domain recommendation~\cite{CERI2012-CrossDomainCF-Survey} mainly focus on {\em relations} between domains, including content-based relations and collaborative filtering based relations. A most recent comprehensive survey on collaborative filtering with additional information~\cite{ACM-CS-2014-YueSHI} focus on different memory-based and model-based methods on exploiting rich side information. Those three surveys consistently emphasize the importance of problem settings or recommendation scenarios such as domains, relations and side information. Note that the types of auxiliary data may also be introduced as an additional dimension for TL-CRAD such as ``where to transfer'' and ``what to transfer'' for different {\em TL settings}, which is a vertical direction with our focus of {\em TL techniques} in this paper.

Besides our focus of TL techniques in this paper, we may also study the representative works in Table~\ref{tbl:summary-TL-CRAD-techniques} from the perspective of TL settings, including four different types of auxiliary data in Table~\ref{tbl:Auxiliary-Data}, and four different sides of auxiliary data, i.e., user side, item side, frontal side (or user-item interaction~\cite{ACM-CS-2014-YueSHI}) and that without overlap. Specifically, CST~\cite{Weike-Pan-AAAI10} is for two-side implicit feedbacks,  CBT~\cite{Bin-Li-IJCAI09} and RMGM~\cite{Bin-Li-ICML09} are for auxiliary explicit feedbacks without overlap, CMF~\cite{CMF-Singh-KDD08} is for item-side content, TCF~\cite{Weike-Pan-IJCAI11} is for frontal-side binary feedbacks, TIF~\cite{AAAI2012-TIF} is for frontal-side uncertain ratings, tagiCoFi~\cite{Yi-Zhen-tagiCoFi-RecSys09} is for frontal-side tags, BMFSI~\cite{Porteous-BPMF-side-infor-AAAI10} is for two-side features, and FM~\cite{TIST12-libFM} is for frontal-side context or user-side content information.

In this paper, we do not include empirical studies of the surveyed representative works, because (i) different TL techniques are usually designed for different recommendation scenarios, (ii) a typical TL technique is usually developed to improve some specific non-TL techniques (e.g., techniques without leveraging auxiliary data or techniques exploiting auxiliary data without explicitly addressing the data difference), and (iii) some TL techniques are designed for different goals though for the same recommendation problem, e.g., TCF~\cite{Weike-Pan-IJCAI11} and iTCF~\cite{IEEE-IS2014-iTCF} are for accuracy and efficiency, respectively. Note that the aforementioned related surveys~\cite{CERI2012-CrossDomainCF-Survey,ICTAI2011-BinLI-CrossDomainCF,ACM-CS-2014-YueSHI} do not include empirical evaluations either.

% =========================================================================================
\subsection{Future Directions}

As a fertile interdisciplinary research area of recommendation and transfer learning, there are various exciting directions worth further exploration in TL-CRAD. In this section, we include several major directions w.r.t. techniques, data, objectives, explanation and security.

{\em Heterogeneous Techniques Ensemble} \; Different transfer learning techniques for CRAD as described in the paper have their own advantages and disadvantages regrading recommendation effectiveness, and learning and prediction efficiency. It is thus natural to design some {\em heterogeneous} knowledge transfer algorithm styles and {\em heterogeneous} knowledge transfer strategies in order to achieve a good balance among flexibility, effectiveness and efficiency. Such heterogeneous TL techniques are expected to be better than a simple combination of existing TL techniques.

{\em Heterogeneous Data Integration} \; An existing transfer learning technique is usually designed for a typical recommendation scenario, while a real recommendation application usually contains more than one types of auxiliary data such as social networks, mobile context and others. Hence, it is very useful to develop a unified framework for {\em heterogeneous} auxiliary data integration. Furthermore, with more and more available data, a scalable and distributed framework for heterogenous data is also needed.

{\em Multi-Objective Recommendation} \; Existing transfer learning techniques in CRAD are mainly for rating prediction and item recommendation, while a real recommendation system requires a multi-objective evaluation such as accuracy, diversity and even serendipity. Hence, it is well motivated to design a more sophisticated objective function with different evaluation metrics when exploiting the auxiliary data.

{\em Explanation and Security} \; Most transfer learning techniques in CRAD are developed for sparisty reduction in the target data. For a real recommendation system, auxiliary data may be taken as a source for explanation generation of the recommended items, and even for robustness against malicious attack.

Practice of leveraging auxiliary data in collaborative recommendation via transfer learning also expands the traditional categorization of recommendation approaches with one more branch, i.e., {\em collaborative recommendation with auxiliary data}, in addition to the two main approaches of {\em content-based recommendation} and {\em collaborative recommendation}. Research or practice of TL-CRAD is also quite interesting in the big data and AI era, especially of the data variability or heterogeneity as commonly known as one of the major properties of today's data.

% =========================================================================================
\subsection{Conclusion}

In this paper, we study collaborative recommendation with auxiliary data (CRAD) from a transfer learning (TL) view. Specifically, we focus on different TL techniques of existing works and answer the question of ``how to transfer'' in transfer learning for CRAD. We first propose a categorization of TL techniques for CRAD, including knowledge transfer algorithm styles and knowledge transfer strategies. We then propose a novel generic knowledge transfer framework for TL-CRAD, and describe some most representative works as instantiations of the proposed framework. Finally, we give some summary discussions and show several exciting directions for further study.

% =========================================================================================
\subsection*{Acknowledgement}
I would like to thank Prof. Qiang Yang for advice and comments.

% =========================================================================================
\bibliographystyle{plain}
\bibliography{paper}

% =========================================================================================
\end{document}